\documentclass{article}
\usepackage{graphicx} % Required for inserting images
\usepackage{amsmath}
\usepackage{color}
\usepackage[abs]{overpic} 

\def\II{\rm II}
\def\babar{{\it BABAR\,}}
\def\bbar{\overline{B}{}^0}

\title{\bf Prospects for the Belle~$\II$ experiment
to further elucidate the KM mechanism and beyond}
\author{Nanae Taniguchi \\
(on behalf of the Belle~$\II$ collaboration) \\
High Energy Accelerator Research Organization  (KEK) \\
Tsukuba, Japan
}
\date{}

\begin{document}

\maketitle

\section{Introduction}
\subsection{Physics at a $B$ factory}

The main approach of flavor physics experiments is to search for 
evidence of new physics via quantum effects.
Since all quark flavors are produced in $e^+ e^-$ collisions, 
one can study possible couplings between new particles and 
any quark flavor, including those of the third generation.
Within the Standard Model (SM), the third generation gives 
rise to $CP$ violation, as described by the 
Kobayashi-Maskawa (KM) mechanism. 
In addition to quarks, an $e^+e^-$ ``$B$ factory''
collider produces a very large number of tau lepton pairs,
which the Belle~$\II$ experiment also studies.

\subsection{SuperKEKB and Belle~$\II$}

SuperKEKB~\cite{superkekb} is the only $e^+e^-$ collider now running at 
the energy of the $\Upsilon$(4S) resonance. The design goal of SuperKEKB 
is to achieve an instantaneous luminosity of 
$10^{35}{\rm cm}^{-2}{s}^{-1}$. To achieve this, the accelerator runs with 
large beam currents and very small beam sizes at the interaction point (IP). 
A comparison of parameters between the previous KEKB accelerator~\cite{kekb_c} 
used for the Belle experiment and the SuperKEKB accelerator used for Belle~II 
is given in Table~\ref{kekb-superkekb}. One notable difference is the beam 
energies, which are less asymmetric at SuperKEKB; this change significantly
improves the beam lifetimes.

\begin{table}[h]
\centering
\begin{tabular}{|l|l|l|}
\hline
~ & KEKB & SuperKEKB \\ \hline
Beam energy (GeV) of $e^- / e^+$ & 8.0/3.5 & 7.0/4.0 \\ \hline
Beam size at IP (mm) & $\sim$ 6 & $\sim$ 0.3\\ \hline
Beam currents (A) & 1.4/1.7 & 2.6/3.6 \\ \hline
Luminosity (${\rm cm}^{-2}{s}^{-1}$) & $2.1 \times 10^{34}$ &  $60 \times 10^{34}$\\
\hline
\end{tabular}
\caption{Main parameters of KEKB and SuperKEKB colliders.}
\label{kekb-superkekb}
\end{table}

The Belle~$\II$ detector consists of multiple components.
It is designed 
to operate at high trigger rates and high background conditions. 
To achieve this, all components have been upgraded from the previous 
Belle detector~\cite{b2tdr}, e.g., with finer segmentation and higher time resolution. 
As the trigger rate at the SuperKEKB design luminosity is expected to reach 30 kHz, 
a pipe-line readout is implemented. 
A new vertex detector consisting of silicon pixels and silicon strips provides 
excellent position resolution for decay vertices of $B$ and $D$ mesons. 
Excellent particle identification is achieved by two newly
developed detectors: a quartz-based ``time-of-propagation'' counter, 
and an aerogel-based ring-imaging Cherenkov counter.

Belle~II began collecting physics data in~2019. 
In 2020 the SuperKEKB collider exceeeded the (instantaneous) luminosity 
record of KEKB of $2\times 10^{34}$, and in 2022 it achieved a world record 
luminosity of $4.7 \times 10^{34}$. As of the summer of 2022, the
integrated luminosity had reached 427~${\rm fb}^{-1}$, which is similar
to that recorded by the \babar\ experiment and about half that
recorded by Belle. The Belle~II experiment is now concluding a 
long shutdown and will begin taking data again in early~2024.
During this shutdown, the second layer of the silicon pixel detector was installed, 
and numerous other improvements were made to the detector and accelerator.

\section{Belle~$\II$ physics program}

Belle (1999--2010) was constructed to confirm the KM model of $CP$ violation. 
This was achieved -- along with the \babar\ experiment at SLAC -- and resulted in
the 2006 Nobel Prize in Physics being awarded to Kobayashi and Maskawa. 
Belle ultimately recorded
almost 1~ab$^{-1}$ of data. Belle~$\II$ is designed to explore beyond the KM model 
and hopefully uncover new physics. The final dataset is expected to be $\sim\!50$~ab$^{-1}$.

\subsection{Search for new physics in mixing}

During the era of Belle and \babar, numerous measurements of the elements of the
Cabibbo–Kobayashi–Maskawa (CKM) mixing matrix were performed. These provide a 
precision test of the unitarity of the CKM matrix, i.e., if the internal 
angles of the CKM ``Unitarity Triangle'' summed up to less than 180$^\circ$, 
or if measurements  of 
the angles determined from tree-amplitude decays and from loop-amplitude 
decays differed, that would indicate physics beyond the~SM.
Belle~II will continue these measurements with much higher precision, 
and it is expected to yield among the most precise determinations 
of all three sides and all three internal angles of this triangle. 

The angle $\phi_1$ provided the first evidence for $CP$ violation in the $B$ system.
This $CP$ violation is caused by interference between a direct decay amplitude and 
a decay amplitude proceeding via mixing. 
The decay $B\to J/\psi\,K_S$ allows one to measure the CKM angle $\phi_1$ with negligible
theoretical uncertainty, as the direct decay amplitude is dominated by a tree diagram and
thus contains essentially a single weak phase. For this decay, there will be a time-dependent 
$CP$ asymmetry between $B^0$ decays and $\bbar$ decays, with an oscillation amplitude given 
by $\sin 2\phi_1$. Decays such as $B\to \phi K^0$ and $B\to \eta' K^0$ should have 
the same weak phase and thus the same dependence upon $\sin 2\phi_1$. However, they 
proceed via $b\to s$ penguin (loop) amplitudes and thus are sensitive to
new particles propagating within internal loops. Such particles would
deviate the amplitude of the time-dependent $CP$ asymmetry from the 
SM prediction ($\sin 2\phi_1$).
With only a small dataset, Belle~$\II$ has measured ${\rm sin}2\phi_1$ for $B\to J/\psi K_S$ 
decays, as listed in Table~\ref{sin2phi1_comparison}. The result is consistent
with the previous Belle measurement; the systematic uncertainties are comparable due to
improvements in the detector performance~\cite{sin2phi1}. In particular, the addition of
pixels to the silicon vertex detector improved the precision of the measurement of 
decay times.

\begin{table}[hbt]
\begin{center}
\begin{tabular}{lcc}
\hline
Experiment & $\sin 2\phi_1$ %\,(J/\psi K^0_S)$ 
                                  & Integrated luminosity \\ 
\hline
Belle~$\II$ & $+ 0.720 \pm 0.062\,({\rm stat}) \pm 0.016\,({\rm syst})$ & 190~fb$^{-1}$ \\
Belle & $+ 0.670 \pm 0.029\,({\rm stat}) \pm 0.013\,({\rm syst})$ & 771~fb$^{-1}$ \\
\hline
\end{tabular}
\caption{\label{sin2phi1_comparison}
Measurements of $\sin 2\phi_1$ by Belle~$\II$ and Belle~\cite{sin2phi1_belle}.}
\end{center}
\end{table}

\begin{figure}[h]
\centering
\includegraphics[width=0.7\textwidth]{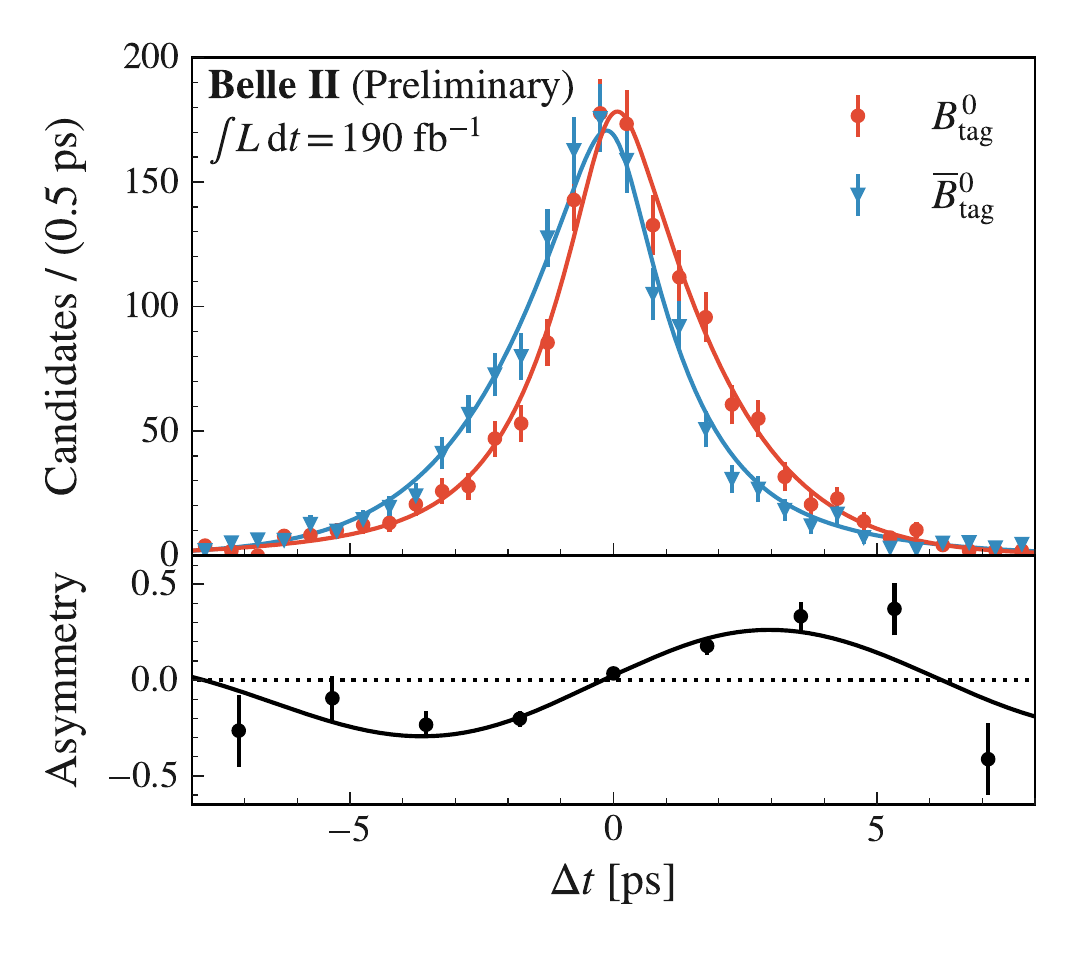}
\caption{\label{fig:fig_1} 
$\Delta t$ distribution for $B\to J/\psi K_S$ decays separated by $B_{\rm tag}$ flavor,
and the \textcolor{black}{raw} asymmetry define as
$[N(B^0_{\rm tag}) - N(\bar{B}^0_{\rm tag})]/[N(B^0_{\rm tag}) + N(\bar{B}^0_{\rm tag})]$.
}
\end{figure}

Figure~\ref{fig:fig_1} shows the $\Delta t$ distribution of $B^0\to J/\psi K_S$ and
$\bbar\to J/\psi K_S$ decays, 
and the measured \textcolor{black}{raw} $CP$ asymmetry, where $\Delta t$ is the
difference between the decay time of the signal $B\to J/\psi K_S$ decay 
(denoted $B^{}_{\rm sig}$) and that of the $B$ meson recoiling against the 
signal decay (denoted $B^{}_{\rm tag}$). The flavor of $B^{}_{\rm sig}$ 
($B^0$ or $\bbar$) is identified by the flavor of~$B_{\rm tag}$.
Further improvements in this measurement are expected as the integrated luminosity increases.
With the full 50~ab$^{-1}$ of data, the angles of the Unitarity Triangle are expected to be 
measured with a precision of $\sim\!1^\circ$, and the sides with a precision of $1\!-\!2\%$.

\subsection{Flavor Changing Neutral Current processes}

Flavor Changing Neutral Current~(FCNC) processes are a good probe for new physics,
as loop diagrams dominate such decays. There are numerous measured observables, and
any deviation from the SM prediction would be interpreted as evidence for new physics.
For radiative or electroweak decays, having a photon or charged leptons in the 
final state results in smaller theoretical uncertainties in the SM prediction.
However, for exclusive decays there are still significant uncertainties due to 
hadronization, e.g., the process $B\to K^{*}$. 
Inclusive measurements have smaller theoretical uncertainties but experimentally
are more challenging.

\begin{figure}[h]
\centering
\includegraphics[width=0.49\textwidth]{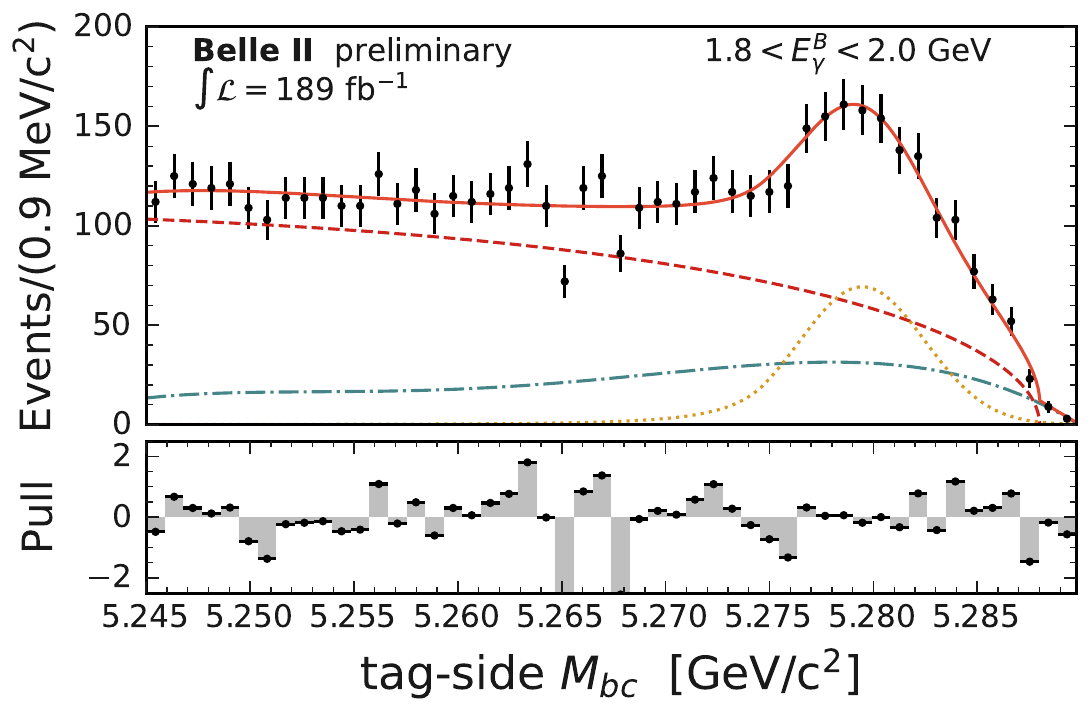}
\includegraphics[width=0.49\textwidth]{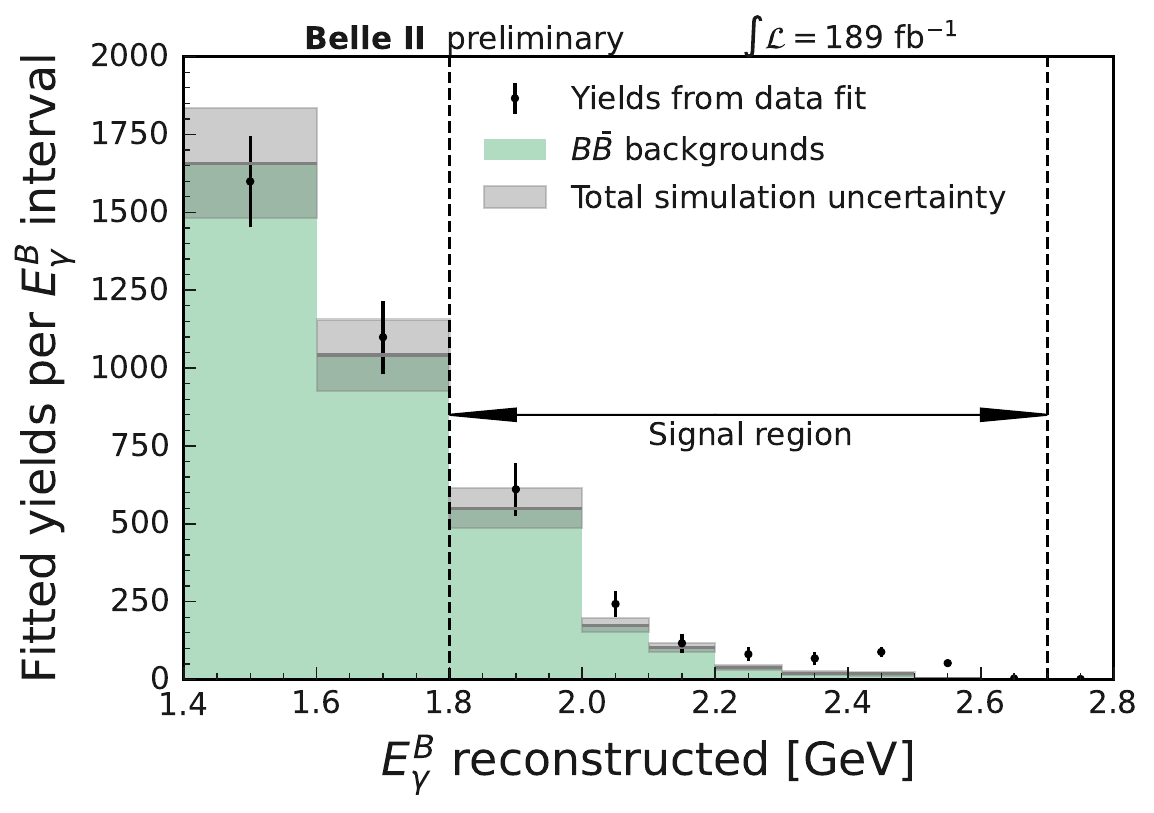}
\caption{\label{fig:fig_2} 
$M^{}_{\rm bc}$ distribution for $B^{}_{\rm tag}$ candidates~(left);
signal yield \textcolor{black}{of $B\bar{B}$ events} as a function of 
photon energy in the $B^{}_{\rm sig}$ rest frame (right).}
\end{figure}

Belle-$\II$ has measured the branching fraction (${\cal B}$) for the inclusive 
radiative decay $B\to X_s\gamma$ using 189~fb$^{-1}$ of data~\cite{xsgamma}. This 
decay proceeds via a $b\to s\gamma$ loop amplitude, and the measurement employs
a ``hadronic tag'' in which $B^{}_{\rm tag}$ (the $B$ meson recoiling against 
$B^{}_{\rm sig}$) 
is fully reconstructed. This reconstruction reduces backgrounds by orders of magnitude, 
but it also greatly reduces the signal efficiency. After all selection criteria, the latter
is only~$\sim\!0.01$\%.
Figure~\ref{fig:fig_2} shows the distribution of the beam-constrained mass ($M^{}_{\rm bc}$) 
for $B^{}_{\rm tag}$, and also the signal yield \textcolor{black}{of $B\bar{B}$ events} 
as a function of photon energy in the $B^{}_{\rm sig}$ rest frame. The observable $M^{}_{\rm bc}$
is defined as 
$\sqrt{E_{\rm beam}^2 - p^2_{\rm tag} }$, 
where $E^{}_{\rm beam}$ is the beam energy and $p_{\rm tag}$ is the reconstructed momentum of 
$B^{}_{\rm tag}$ in the $e^+e^-$ center-of-mass frame.
The result, 
\textcolor{black}{${\cal B}$($B\to X_s\gamma) = 3.54 \pm 0.78~({\rm stat.} ) \pm 0.83~({\rm syst.})$ for $E^{B}_{\gamma} > 1.8~{\rm GeV}$}, 
is consistent with the SM prediction; the precision is 
similar to that obtained by \babar\ using 210~fb$^{-1}$ of data. 
Both measurements are dominated by systematic uncertainties. 
In the future, Belle $\II$ expects to reduce the overall uncertainty from 5\% to~3\%. 
\textcolor{black}{
The dominant systematic error in the lepton-tag method comes from a fake signal of photon due to neutral hadrons~\cite{b2book}.
}

The electroweak penguin decay $B\to K^{(*)}\ell^+\ell^-\,(\ell = e,\mu)$ is suppressed more 
than the radiative decay $B\to X_s\gamma$. Nonetheless, Belle~II measured this decay
with a data set of only 189~fb$^{-1}$~\cite{LFU}. As the electron and muon channels have 
similar detection efficiencies, the ratio
$R_{K^{(*)}}\equiv {\cal B}(B\to K^{(*)}\mu^+\mu^-)/{\cal B}(B\to K^{(*)}e^+e^-)$ 
provides \textcolor{black}{an important} test of lepton flavor universality. 

The results are:
\begin{eqnarray*}
{\cal B}(B\to K^{*} \mu^+\mu^-) & = & (1.19 \pm 0.31\,^{+0.08}_{-0.07}) \times 10^{-6} \\
{\cal B}(B\to K^* e^+e^-)       & = & (1.42 \pm 0.48 \pm 0.09) \times 10^{-6} 
\end{eqnarray*}
\textcolor{black}{
These values imply a ratio $0.83 \pm 0.36$, which is consistent with the theory expectation~\cite{RK_SM}.
}

With more data, these measurements should significantly improve.
Scaling these uncertainties by luminosity, 
one obtains $\sim\!3\%$ precision for $50~{\rm ab}^{-1}$ of data
\textcolor{black}{(for $q^2={M_{\ell\ell}}^2$ bin [1-6]~${\rm GeV}^2/c^4$ )}~\cite{{RK_prospect}}.

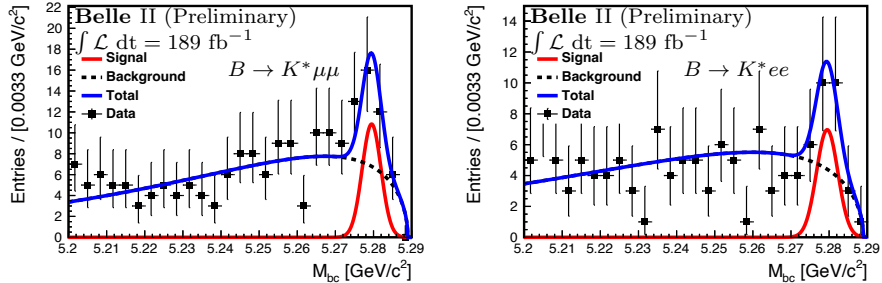
\begin{figure}[h]
\begin{overpic}[width=0.49\textwidth]{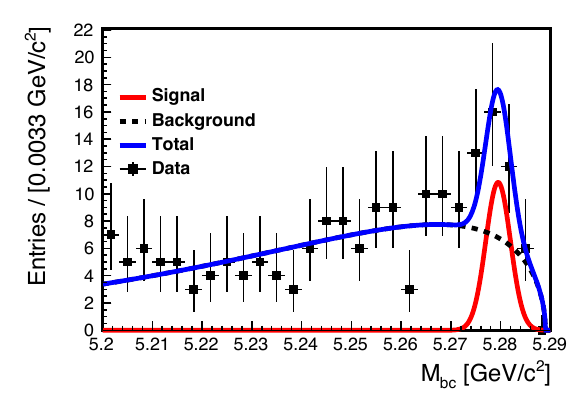}
 \put (32,106) {\footnotesize {{\bf{Belle~$\II$}} (Preliminary)}}
 \put (32,96.5) {\footnotesize {$\int \mathcal{L}~\rm{dt}=189~\rm{fb}^{-1}$}}
 \put (90,86) {\footnotesize $B\to K^*\mu\mu$}
\end{overpic}
\begin{overpic}[width=0.49\textwidth]{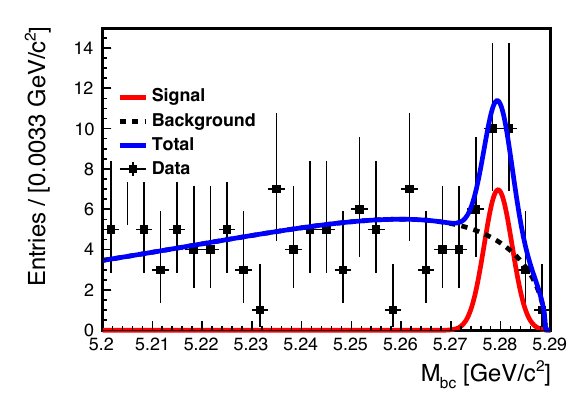}
 \put (32,106) {\footnotesize {{\bf{Belle~$\II$}} (Preliminary)}}
 \put (32,96.5) {\footnotesize {$\int \mathcal{L}~\rm{dt}=189~\rm{fb}^{-1}$}}
 \put (90,86) {\footnotesize $B\to K^* ee$} 
\end{overpic}
\caption{\label{fig:fig_3} 
$M_{\rm bc}$ distributions for $B\to K^*\mu^+\mu^-$ (left) and $B\to K^{*}e^{+}e^{-}$ (right).}
\end{figure}

\begin{figure}[h]
\centering
\includegraphics[width=1\textwidth]{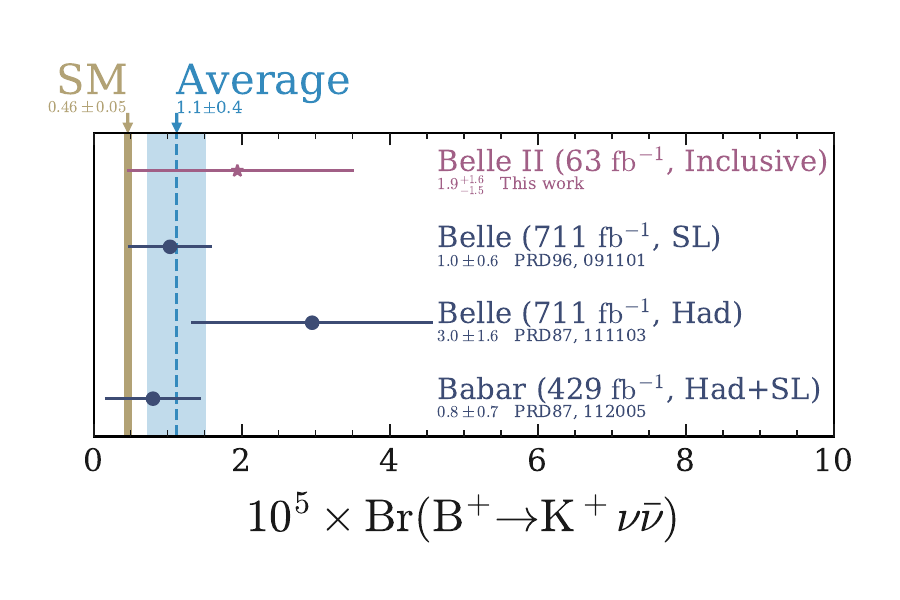}
\vskip-0.20in
\caption{\label{fig:fig_4}
Branching fraction for $B^+\to K^+\nu\bar{\nu}$ measured by Belle~$\II$ and 
Belle~\cite{knn_14,knn_15,knn_16}. Also shown is the SM prediction~\cite{knn_17}.
}
\end{figure}

The electroweak penguin decay $B\to K^{(*)}\nu\bar{\nu}$ has not yet been observed.
This decay is especially challenging to reconstruct as it has two neutrinos in the 
final state. Belle~$\II$ is the only running experiment able to search for this decay.
Using only 63~fb$^{-1}$ of data, Belle~II searched for $B^+\to K^{+}\nu\bar{\nu}$ 
using a new analysis method~\cite{knn}. The signal $K^+$ is identified as the 
charged track with highest transverse momentum that satisfies particle identification 
criteria. All remaining tracks and energy clusters are associated with the other $B$ 
meson in an event ($B^{}_{\rm tag}$). This results in high reconstruction efficiency but
also high background levels. The latter is reduced by using a  
boosted decision tree classifier that is trained to identify distinctive 
characteristic of signal events. No signal is observed, and an upper limit on 
the branching fraction is obtained:
\begin{eqnarray*}
{\cal B}(B\to K^+ \nu\bar{\nu}) & < & 4.1 \times 10^{-5} \hskip0.20in (90\%\ {\rm C.L.}) 
\end{eqnarray*}
This precision is similar to that obtained by Belle using 711~fb$^{-1}$ of data and 
a hadronic tagging method. All results are summarized in Figure~\ref{fig:fig_4}.

\subsection{Tau physics prospects}

In addition to $B\overline{B}$ events, Belle~$\II$ will collect a large sample of $\tau^+\tau^-$
events. This sample will provide a rich physics programs of high precision measurements.
For example, Belle~II can search for lepton flavor violation~(LFV) in such modes as 
$\tau^+\to\mu^+\gamma$, $\tau^+\to \ell^+\ell^-\ell^+\,(\ell=\mu,e)$, etc. There are 
more than 40 such LFV $\tau^+$ decays, many of which can only be reconstructed at 
an $e^+e^-$ experiment such as Belle~II.

\subsection{Dark sector prospects}

Several studies of the dark sector are active at Belle~$\II$.
Dark matter may interact with SM particles through various ``portal'' interactions.
Belle~$\II$ can search for dark matter having a mass in the range 100~MeV/$c^2$ to 
a few~GeV/$c^2$. Many decay signatures consist of only a single photon or a single 
track in the detector, and specialized triggers have been developed for Belle~II 
to record such topologically unusual events.

\section{Summary}

In summary, a new $e^+e^-$ collider (SuperKEKB) has been constructed and commissioned
at KEK, and a new state-of-the-art detector (Belle~II) has begun taking data and 
producing physics results. This collider is the most recent in a series that began 
with TRISTAN and continued with KEKB. The SuperKEKB collider has achieved a world 
record instantaneous luminosity of $4.7\times 10^{34}~{\rm cm}^{-2}s^{-1}$, and 
Belle~II has already recorded an integrated luminosity of several hundred fb$^{-1}$, 
equivalent to the \babar\ dataset. This is the start of a new era of measurements 
at a ``Super $B$ factory,'' with the goal of uncovering new physics beyond 
the~SM.

\end{document}